\documentclass[%
 preprint,
 amsmath,amssymb,
 aps,
]{revtex4-2}

\usepackage{graphicx}
\usepackage{dcolumn}
\usepackage{bm}
\usepackage{multirow}
\usepackage{nicematrix}
\usepackage{diagbox}

\usepackage{xcolor}

\begin{document}

\preprint{APS/123-QED}

\title{Data-driven low-dimensional model of a sedimenting flexible fiber}

\author{Andrew J. Fox}
\author{Michael D. Graham}%
 \email{mdgraham@wisc.edu.}
\affiliation{ 
Department of Chemical and Biological Engineering,\\
University of Wisconsin-Madison,\\
Madison, WI 53706, USA
}%

\date{\today}

\begin{abstract}
The dynamics of flexible filaments entrained in flow, important for understanding many biological and industrial processes, are computationally expensive to model with full-physics simulations.
This work describes a data-driven technique to create high-fidelity low-dimensional models of flexible fiber dynamics using machine learning; the technique is applied to sedimentation in a quiescent, viscous Newtonian fluid, using results from detailed simulations as the data set.
The approach combines an autoencoder neural network architecture to learn a low-dimensional latent representation of the filament shape, with a neural ODE that learns the evolution of the particle in the latent state.  
The model was designed to model filaments of varying flexibility, characterized by an elasto-gravitational number $\mathcal{B}$, and was trained on a data set containing the evolution of fibers beginning at set angles of inclination. For the range of $\mathcal{B}$ considered here (100-10000), the filament shape dynamics can be represented with high accuracy with only four degrees of freedom, in contrast to the 93 present in the original bead-spring model used to generate the dynamic trajectories.  
We predict the evolution of fibers set at arbitrary angles and demonstrate that our data-driven model can accurately forecast the evolution of a fiber at both trained and untrained elasto-gravitational numbers.
\end{abstract}

\maketitle

\section{Introduction}

The dynamics of flexible filaments in flow are an active area of interest, with applications in many biological and industrial systems. \cite{DuRoure2019, Yu2023}
Biopolymers, such as DNA, can behave like elastic fibers suspended in flow, and understanding their dynamics is important for designing and understanding various medical procedures. \cite{Shelley2016}
In paper manufacturing, the development and processing of fibers from wood pulp is vital for creating product at given specifications and for optimizing production to reduce waste. \cite{Lundell2011}
In particular, the relationship and competition between hydrodynamic forces acting on the particle and elastic forces acting along the fiber can produce a range of interesting behavior. \cite{Graham2011, Yu2021, Yu2022}

In recent years, modeling the sedimentation dynamics of flexible filaments has been explored through experiments and numerical analyses.
\citet{Xu1994} investigated the sedimentation of a flexible slender particle with small deformation in quiescent viscous fluid.
They performed a theoretical analysis of a slender body undergoing small amplitudes of deformation using the Euler-Bernoulli beam theory, showing a "U-shape" conformation at long settling times.
\citet{Schlagberger2005} furthered this analysis by performing computational simulations of the system.
Using numerical computations of a settling filament of a bead-spring-model, they  expanded this analysis to understand the dynamics at higher amplitudes of deformation.
\citet{Lagomarsino2005} continued the use of bead-spring models to investigate the sedimentation of highly flexible filaments.
They analyzed very flexible filaments with a bead-spring model, revealing a stable "W-shape" confirmation for sufficiently high degrees of flexibility.
\citet{Li2013} used analytical and numerical analysis to understand the settling dynamics of an elastic fiber when oriented along its major axis.
They applied a slender body theory to highly flexible filaments, identifying a buckling instability when sedimenting in this configuration.
\citet{Shojaei2015} investigated the rotational dynamics of a flexible filament settling in quiescent flow.
Using numerical simulations of a slender beam model, they identified different rotational behaviors of filaments at different degrees of elasticity.
\citet{Delmotte2015} developed a generalized formulation of bead spring models with which to investigate the dynamics of flexible fibers outside of Stokes flow.
Their approach allowed them to explore the dynamics of elastic filaments at finite Reynolds numbers.
\citet{Marchetti2018} performed both experiments and numerical computations to analyze the settling behavior of flexible filaments.
Using a bead-spring model, they successfully demonstrated a quantitative agreement between these studies, identifying distinct regimes of particle deformation.
\citet{Cunha2022} compared the dynamics of non-Brownian and Brownian flexible filaments settling in quiescent fluid.
Using experiments and numerical simulations, they demonstrated good agreement between the observed and computed dynamics for both filament regimes.

As the interplay between the hydrodynamic and elastic forces in these studies can be complex and the entire particle structure must be adequately captured, full physics simulations of fluid dynamics can be computationally expensive \citep{Cheng2023, Ebrahimi2022}.
One potential solution to model fluid and particle mechanics without long computational times is to develop low-dimensional, data-driven models\citep{Linot2022, Linot2023, Vinuesa2022}. 
These models, created though machine learning, can capture the evolution of a dynamical system with significantly fewer dimensions than full physics simulations, with minimal loss of prediction accuracy\citep{Fox2023, Linot2020, Omata2019}. 
These models use latent representations of the full-field data to learn a parameterization of the manifold on which the dynamics of a system exist, primarily through the use of undercomplete autoencoder neural networks\citep{Zeng2023a, Perez2022, Brunton2020}. 
By combining this with a time-integrating neural network, either in discrete or continuous time, the full dynamics of the system can be captured with much less computational expense\citep{Zeng2023b, Srinivasan2019}. 

Here, we will develop data-driven models to forecast the dynamics of a flexible fiber settling  in a quiescent Newtonian fluid in the Stokes flow regime. To our knowledge, such modeling approaches have not been applied to deformable-particle dynamics problems.
Specifically, we will model the trajectory and shape evolution of a filament, beginning lineally-aligned at rest at an arbitrary angle of initial orientation, and evolving until the fiber reaches a terminal configuration.
We will train autoencoder neural networks to learn a low-dimensional representation of the shape of a fiber, then use a time-integrating neural network to learn both the evolution of the filament shape and the position; these will be combined to produce a single low-dimensional model of the fiber dynamics.
We will demonstrate that the fiber structure and dynamics can be captured at significantly lower dimension than the full-field state.
We will demonstrate that our model can forecast the evolution of fibers at arbitrary angles of initial inclination and can predict the dynamics of fiber shape and position at a range of elasto-gravitational numbers with minimal error.

\section{Formulation}

\subsection{Physics-based model}

\begin{figure}
\centering
\includegraphics[width=0.5\textwidth]{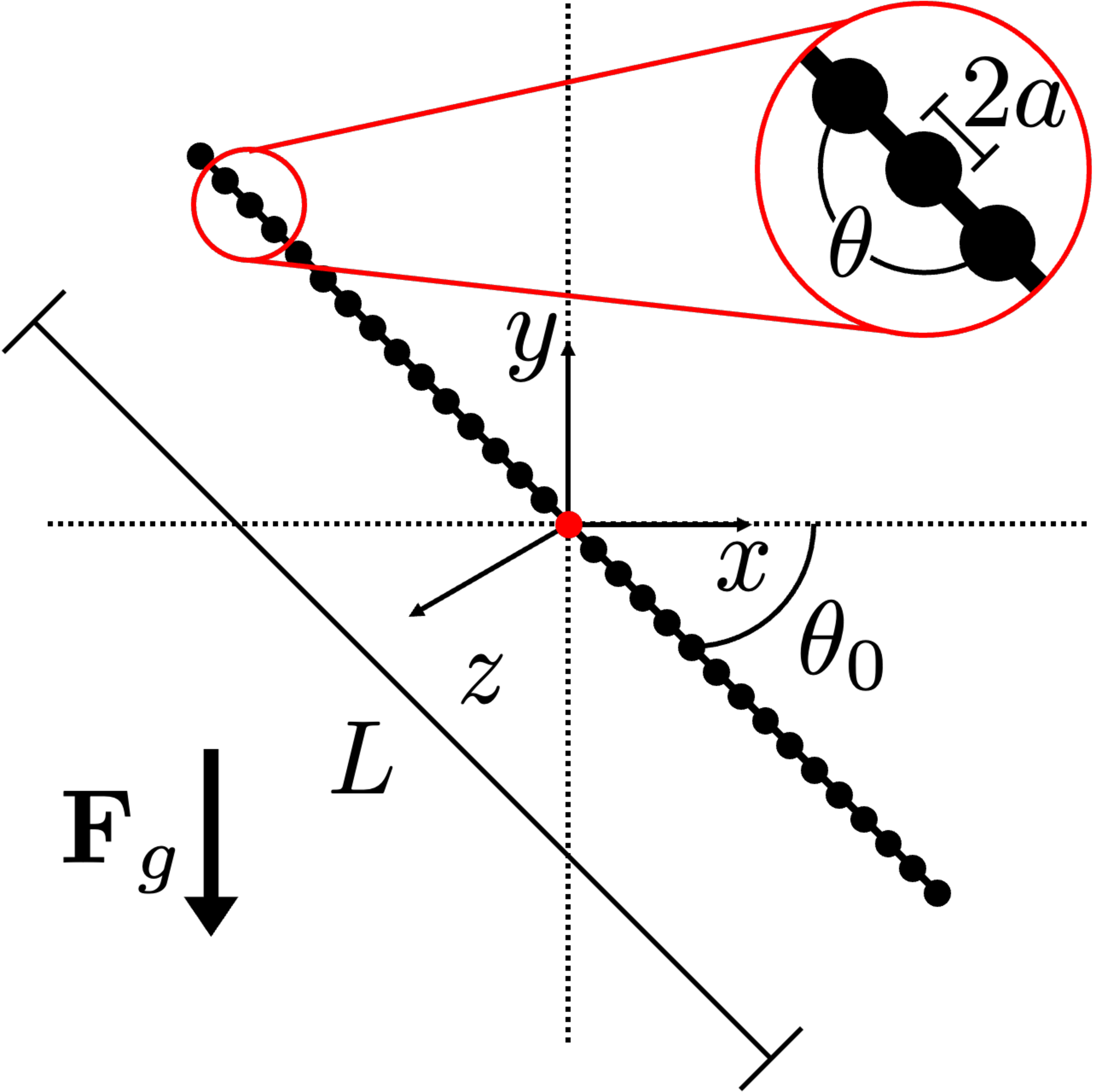}
\caption{A flexible filament of length $L$ settling under an external force $\mathbf{F}_g$ in a quiescent Newtonian fluid. The filament begins at an arbitrary initial inclination, $\theta_0$, relative to the $x$-axis, and evolves until it reaches a terminal shape. The fiber is modeled as a series of $N$ beads of radius $a$ connected by springs, with the center bead, which has position $\mathbf{c}(t)$, shown in red.}
\label{ProblemDiagram}
\end{figure}

Consider a flexible filament of length $L$ entrained in a quiescent, viscous Newtonian fluid of viscosity $\mu$, driven by a gravitational force, $\mathbf{F}_g$, and experiencing a viscous drag, $\mathbf{F}_{drag}$, dependent on the instantaneous fiber shape.
The fiber begins aligned in the $xy$-plane at rest, set at an arbitrary initial angle of inclination, $\theta_0$, relative to the $x$-axis.
The filament is allowed to settle under an external force (typically considered to be gravity), whereupon the elasticity of the fiber induces a deformation of its shape, trending towards a ``U-shape'' bending\citep{Xu1994}.
At long times, the shape of the fiber will approach a terminal ``U-shape'', with the particular shape dependent on the balance between gravitational and elastic forces, given by an elasto-gravitational number $\mathcal{B}$\citep{Marchetti2018}.
A diagram of the problem is shown in Figure \ref{ProblemDiagram}.

To model this system, we apply a bead-spring model, wherein the fiber is discretized into a series of $N = 31$ spherical beads of radius $a$, shown by \citet{Marchetti2018} to adequately capture the range of rotational dynamics of interest in this work.
The spheres are connected by massless springs of equilibrium length, $l_0 = 2a$ with an equilibrium angle formed by three beads of $\theta = \pi$.
The fluid is modeled as a viscous Newtonian fluid of viscosity $\mu$ governed by the Stokes equations, with fluid and particle inertia neglected.
The evolution of the filament shape and position is governed by a force balance, with each bead experiencing gravitational and viscous drag forces, and the beads interacting through multi-body hydrodynamic interactions and elastic forces.
The velocity of bead $\alpha$, given by $\dot{\mathbf{X}}^\alpha$, at position $\mathbf{X}^\alpha$ and interacting with beads $\beta$ at position $\mathbf{X}^\beta$ is given by \citep{Marchetti2018}
\begin{equation}
\dot{X}_i^\alpha=\sum_\beta \mathcal{M}_{i j}^{\alpha \beta}\left(F^\beta_j-\frac{\partial \mathcal{U}}{\partial X_j^\beta}\right).
\end{equation}
Here $\mathcal{M}_{i j}^{\alpha \beta}$ is the position-dependent $N$-body mobility tensor, which accounts for hydrodynamic interactions between spheres $\alpha$ and $\beta$ \cite{Graham.2018.https://doi.org/10.1017/9781139175876}, and 
$\mathbf{F}^\beta$ is the external force exerted on each bead by gravity, which in a viscous fluid is balanced by Stokes drag and is given by $\mathbf{F}^\beta = 6 \pi \mu a \mathbf{U}_s$, where $\mathbf{U}_s$ is the Stokes settling velocity.
The elastic potential, $\mathcal{U}$, stems from a discrete form of the wormlike chain model, written as \citep{Schlagberger2005}
\begin{equation}
\mathcal{U}=\sum_\gamma\left[a S\left(\frac{\mathbf{X}^{\gamma, \gamma+1}}{2 a}-1\right)^2+\frac{B}{2 a}\left(1-\cos \theta^{\gamma, \gamma+1}\right)\right].
\end{equation}
For an isotropic elastic cylinder, the stretching modulus, $S$, and bending modulus $B$, depend solely on the cylinder radius and Youngs modulus, $E$, such that $S = E \pi a^2$ and $B = E \pi a^4 / 4$, respectively; this formulation allows the stretching and bending of the filaments to be governed by a single physical parameter and has been shown to accurately capture the dynamics of flexible fibers\citep{Marchetti2018}. 
The geometric dependence of the elastic potential is given by the distance between neighboring beads $\gamma$ and $\gamma + 1$, $\mathbf{X}^{\gamma , \gamma + 1} = \mathbf{X}^{\gamma +1} - \mathbf{X}^\gamma$, and the angle between neighboring springs $\mathbf{X}^{\gamma , \gamma + 1}$ and $\mathbf{X}^{\gamma , \gamma - 1}$, $\theta^{\gamma , \gamma + 1}$.
The inter-particle mobility tensor is selected as the Rotne-Prager-Yamakawa tensor, given for $\alpha\neq\beta$  by \citep{Marchetti2018}

\begin{equation}
\mathcal{M}_{i j}^{\alpha \beta}=\frac{1}{6 \pi \mu a}\left\{\frac{3}{4}\left[\frac{\delta_{i j}}{\frac{X^{\alpha, \beta}}{a}}+\frac{\frac{X_i^{\alpha, \beta} X_j^{\alpha, \beta}}{a^2}}{\left(\frac{X^{\alpha, \beta}}{a}\right)^3}\right] \right. + \left. \frac{3}{2}\left[\frac{\delta_{i j}}{3\left(\frac{X^{\alpha, \beta}}{a}\right)^3}-\frac{\frac{X_i^{\alpha, \beta} X_j^{\alpha, \beta}}{a^2}}{\left(\frac{X^{\alpha, \beta}}{a}\right)^5}\right]\right\}.
\end{equation}

The chosen tensor accounts for hydrodynamic interactions between particles up to order $O\left(\frac{a}{X^{\alpha, b}}\right)^3$, where $X^{\alpha , \beta} = |\mathbf{X}^{\alpha , \beta}| = |\mathbf{X}^\beta - \mathbf{X}^\alpha|$ is the distance between beads $\alpha$ and $\beta$.
A full Rotne-Prager-Yamakawa tensor, which provides a separate formulation for $X^{\alpha , \beta} < 2a$ and is positive definite for all particle configurations, has been observed by \citet{Marchetti2018} to not significantly change the filament dynamics in the parameter range of interest in this study, as the spheres in the bead-spring model do not significantly overlap. The self-mobility $\mathcal{M}_{i j}^{\alpha \alpha}$ is given by the Stokes mobility\citep{Marchetti2018}.
The bead velocity equation can be nondimensionalized by the length scale $a$ and the force scale $6  \pi \mu a U_s$, resulting in

\begin{equation}
\dot{\hat{X}}_i^\alpha=\sum_\beta \hat{\mathcal{M}}_{i j}^{\alpha \beta}\left(\hat{F}_j^\beta-\mathcal{E} \frac{\partial \hat{\mathcal{U}}}{\partial X_j^\beta}\right),
\end{equation}
where $\hat{\mathbf{X}}(t)$ is the dimensionless bead positions and $\hat{F}_j^\beta$ is a downward unit vector.
Now a dimensionless modulus $\mathcal{E}=\frac{E \pi a^2}{6 \pi \mu a U_S}$ appears as the only parameter in the formulation. This can be related to the elasto-gravitational number by $\mathcal{B}=32 \Delta \rho g \ell^3 / E a^2 \equiv 24 (a/l)^{-3} / \mathcal{E}$; here, the volume of the filament is calculated via a bent cylinder encompassing the bead-spring system, the use of which has been shown to successfully match the behaviors of flexible fibers in experiments\citep{Marchetti2018}.

In this paper, we investigated the settling dynamics of settling fibers with elasto-gravitational numbers in the range $100 \leq \mathcal{B} \leq 10000$.
At each $\mathcal{B}$, we calculated the trajectory of filaments beginning at 23 initial orientations, evenly spaced in the range $0 \leq\theta_0 \leq 11 \pi / 24$.
Integration of the positions of the bead-spring system was performed using the ``dopri'' integrator of the "ode" solver in Python, an explicit Runge-Kutta method of order (4)5, in a method adapted from \citet{Marchetti2018}.
The shape and position of each filament was allowed to evolve until the fiber reached a terminal shape, dependent on $\mathcal{B}$, taking a total settling time $T = T(\mathcal{B}, \theta_0)$.

\begin{figure*}
\centering
\includegraphics[width=1\textwidth]{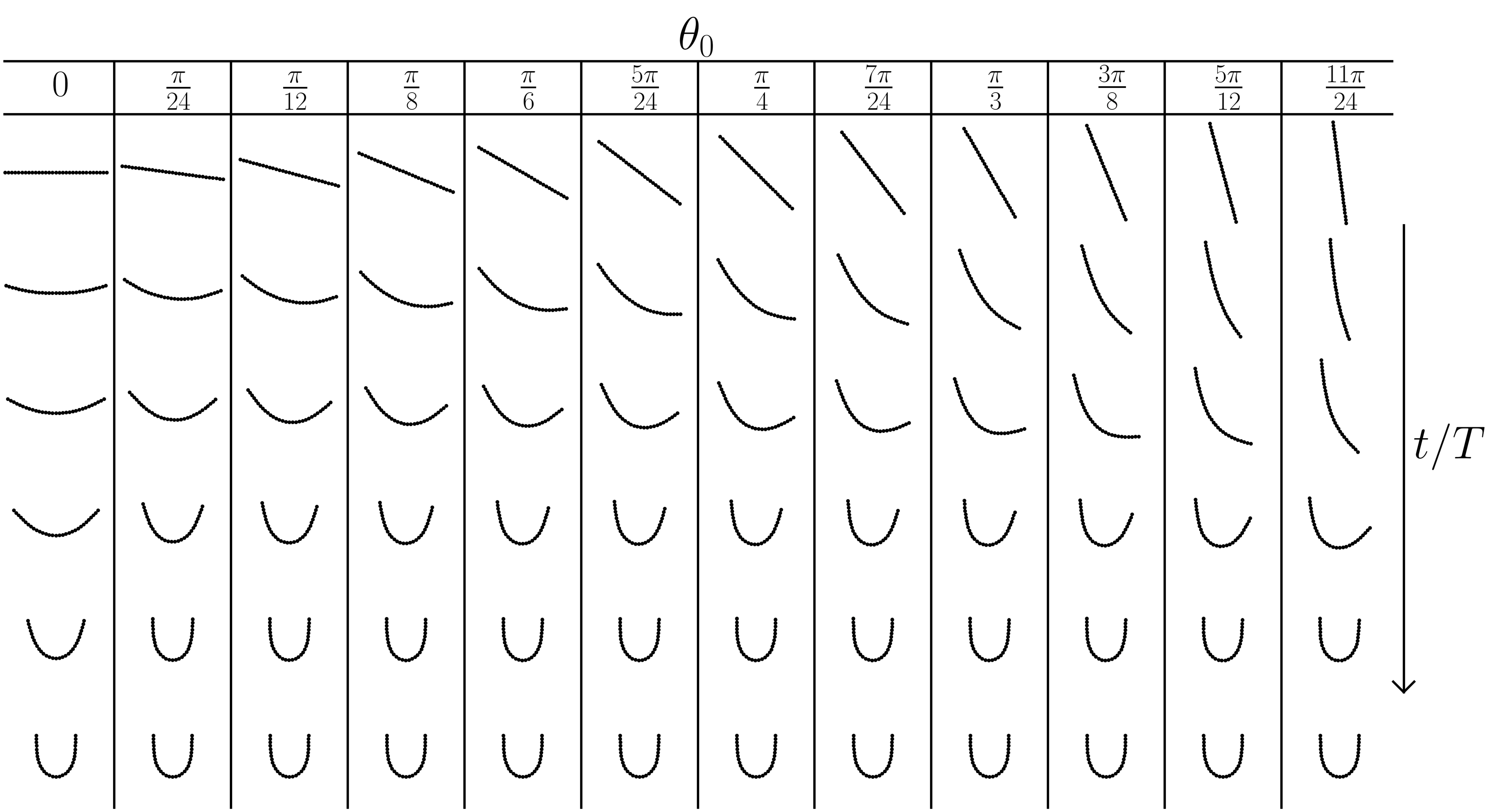}
\caption{The evolution of the shape of a filament settling in a quiescent Newtonian fluid from an initial orientation to a common terminal shape at $\mathcal{B} = 1000$ for all initial angles of orientation within the training data set.}
\label{Shape_B=1000}
\end{figure*}

The evolution of the shape of filaments of $\mathcal{B} = 1000$ at twelve initial orientations is shown in Figure \ref{Shape_B=1000}.
In each trajectory, the filament bends into a ``U-shape'' configuration and rotates into a position such that the shape is symmetric in the $x$-direction.
The terminal shape is identical for all initial orientations for filaments of a given $\mathcal{B}$.

\begin{figure}
\centering
\includegraphics[width=0.5\textwidth]{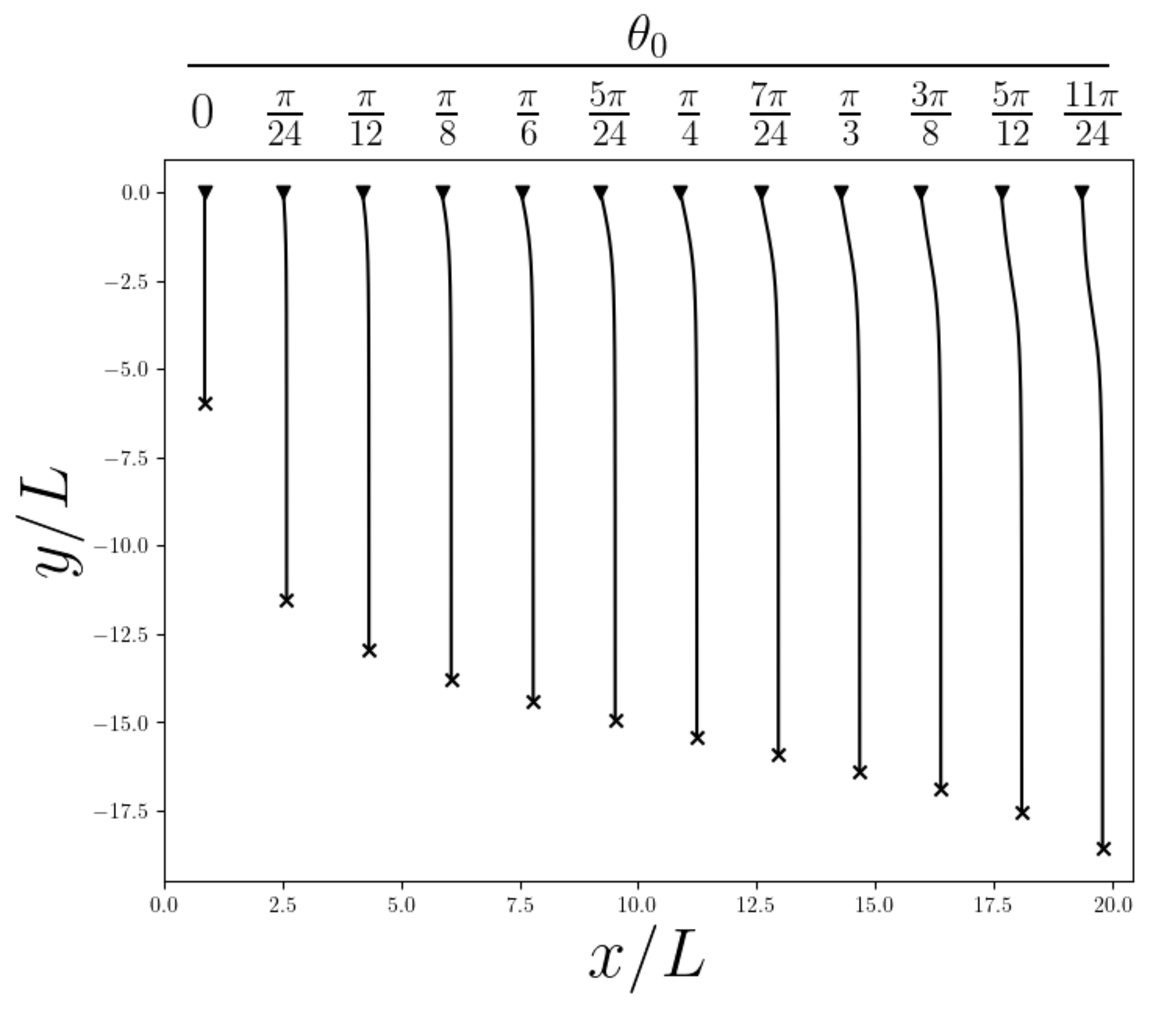}
\caption{The trajectories  $\mathbf{c}(t)$ of the center bead of a filament settling in a quiescent Newtonian fluid at $\mathcal{B} = 1000$ for all initial angles of inclination within the training data set. The initial positions are denoted by symbol “$\blacktriangledown$” and the terminal positions are denoted by the symbol “$\times$”.}
\label{Trajectory_B=1000}
\end{figure}

Shown in Figure \ref{Trajectory_B=1000} is the trajectory of the center bead of a modeled filament, beginning at twelve different initial angles of inclination.
The initial positions of the filaments are arbitrary, and are shown here set at $y = 0$ and offset in the $x$-direction to differentiate between trajectories.
The total distance traveled by each filament depends on the initial orientation, as the time required to reach the terminal shape depends on the bending and rotation required.
The settling time and positional translation increases as the initial angle of orientation increases, as further time is necessary to reorient into a symmetric shape.

\begin{figure*}
\centering
\includegraphics[width=0.6\textwidth]{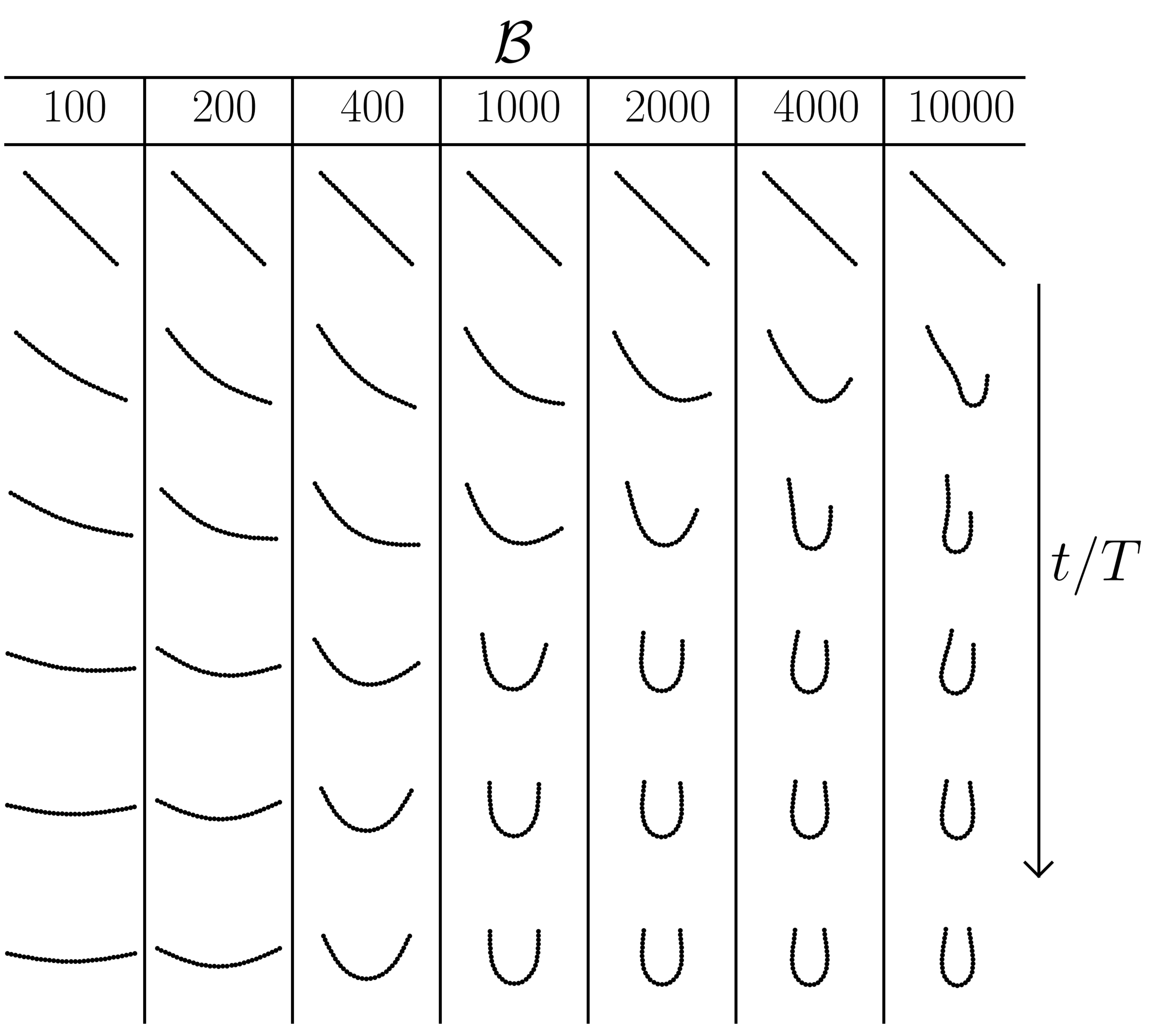}
\caption{The evolution of the shape of a filament settling  in a quiescent Newtonian fluid from a common initial angle of orientation of $\pi/4$ to a terminal shape for all $\mathcal{B}$ within the training data set.}
\label{Shape_B=in_Theta=12}
\end{figure*}

In the range of $\mathcal{B}$ investigated in this study, the flexible filament undergoes three distinct regimes of rotational dynamics.
Examples of these dynamics are shown in Figure \ref{Shape_B=in_Theta=12}, with shape evolution shown for $\theta_0 = \pi/4$ at seven elasto-gravitational numbers within the scope of this investigation.
At low $\mathcal{B}$, the shape dynamics are primarily rotational, with the stiff fiber bending only slightly in the terminal shape.
At moderate $\mathcal{B}$, the rotational dynamics exist in a bending regime, where the left and right sides of the filament bend toward each other to reach the terminal shape and only slight rotation occurs.
For high $\mathcal{B}$, the filament undergoes a ``snaking'' behavior, where the lower side of the fiber bends over itself and slides along the higher side to reach its final shape.
The terminal shape, as shown, strongly and solely depends on the given elasto-gravitational number, with higher $\mathcal{B}$ producing more pronounced "U-shapes" which are identical, to an arbitrary degree of precision, within each $\mathcal{B}$.

\subsection{Data-driven model}

Using the data from the bead-spring model, data-driven models were trained to forecast the evolution of the shape and position of a filament beginning at an arbitrary initial angle of orientation.
A total of 227 trajectories were generated with $100 \leq \mathcal{B} \leq 10000$ and $0 \leq \theta \leq 22 \pi/48$.
The calculated trajectories were separated into three data sets: the training data, test data set A, and test data set B. A breakdown of the distribution of the trajectories into each data set is shown in Table \ref{DataDistribution}.
The training data (green on Table \ref{DataDistribution}) consisted of trajectories of filaments with elasto-gravitational numbers $\mathcal{B} = [100, 200, 400, 1000, 2000, 4000, 100000]$ at twelve initial angles of inclination, evenly space in the range  $0 \leq \theta_0 \leq 22 \pi / 48$.
Test data set A (yellow) consisted of trajectories of fibers at the same elasto-gravitational numbers as the training data set at eleven untrained initial angles of inclination, evenly spaced in the range $\pi/48 \leq \theta_0 \leq 21 \pi/48$.
Test data set A allows for evaluating the ability of the data-driven models to interpolate the dynamics of the settling filaments to different  initial angles within trained $\mathcal{B}$.
Test data set B (red) consisted of filament trajectories at elasto-gravitational numbers $\mathcal{B} = [150, 300, 700, 1500, 3000, 7000]$.
Test data set B allows for evaluating the ability of the data-driven models to interpolate the dynamics of the settling filaments to different initial angles and elasto-gravitational numbers.

\begin{table*}
\begin{NiceTabular}{|c|c|c|c|c|c|c|c|c|c|c|c|c|c|c|c|c|c|c|c|c|c|c|c|}
\hline
\diagbox{$\mathcal{B}$}{$48 \theta_0 / \pi$} & \hspace{0.032in}0\hspace{0.032in} & \hspace{0.032in}1\hspace{0.032in} & \hspace{0.032in}2\hspace{0.032in} & \hspace{0.032in}3\hspace{0.032in} & \hspace{0.032in}4\hspace{0.032in} & \hspace{0.032in}5\hspace{0.032in} & \hspace{0.032in}6\hspace{0.032in} & \hspace{0.032in}7\hspace{0.032in} & \hspace{0.032in}8\hspace{0.032in} & \hspace{0.032in}9\hspace{0.032in} & 10 & 11 & 12 & 13 & 14 & 15 & 16 & 17 & 18 & 19 & 20 & 21 & 22 \\ \hline
100   & \Block[fill=green]{}{} & \Block[fill=yellow]{}{} & \Block[fill=green]{}{} & \Block[fill=yellow]{}{} & \Block[fill=green]{}{} & \Block[fill=yellow]{}{} & \Block[fill=green]{}{} & \Block[fill=yellow]{}{} & \Block[fill=green]{}{} & \Block[fill=yellow]{}{} & \Block[fill=green]{}{} & \Block[fill=yellow]{}{} & \Block[fill=green]{}{} & \Block[fill=yellow]{}{} & \Block[fill=green]{}{} & \Block[fill=yellow]{}{} & \Block[fill=green]{}{} & \Block[fill=yellow]{}{} & \Block[fill=green]{}{} & \Block[fill=yellow]{}{} & \Block[fill=green]{}{} & \Block[fill=yellow]{}{} & \Block[fill=green]{}{} \\ \hline
150   &   & \Block[fill=red]{}{} &   & \Block[fill=red]{}{} &   & \Block[fill=red]{}{} &   & \Block[fill=red]{}{} &   & \Block[fill=red]{}{} &    & \Block[fill=red]{}{} &    & \Block[fill=red]{}{} &    & \Block[fill=red]{}{} &    & \Block[fill=red]{}{} &    & \Block[fill=red]{}{} &    & \Block[fill=red]{}{} &    \\ \hline
200   & \Block[fill=green]{}{} & \Block[fill=yellow]{}{} & \Block[fill=green]{}{} & \Block[fill=yellow]{}{} & \Block[fill=green]{}{} & \Block[fill=yellow]{}{} & \Block[fill=green]{}{} & \Block[fill=yellow]{}{} & \Block[fill=green]{}{} & \Block[fill=yellow]{}{} & \Block[fill=green]{}{} & \Block[fill=yellow]{}{} & \Block[fill=green]{}{} & \Block[fill=yellow]{}{} & \Block[fill=green]{}{} & \Block[fill=yellow]{}{} & \Block[fill=green]{}{} & \Block[fill=yellow]{}{} & \Block[fill=green]{}{} & \Block[fill=yellow]{}{} & \Block[fill=green]{}{} & \Block[fill=yellow]{}{} & \Block[fill=green]{}{} \\ \hline
300   &   & \Block[fill=red]{}{} &   & \Block[fill=red]{}{} &   & \Block[fill=red]{}{} &   & \Block[fill=red]{}{} &   & \Block[fill=red]{}{} &    & \Block[fill=red]{}{} &    & \Block[fill=red]{}{} &    & \Block[fill=red]{}{} &    & \Block[fill=red]{}{} &    & \Block[fill=red]{}{} &    & \Block[fill=red]{}{} &    \\ \hline
400   & \Block[fill=green]{}{} & \Block[fill=yellow]{}{} & \Block[fill=green]{}{} & \Block[fill=yellow]{}{} & \Block[fill=green]{}{} & \Block[fill=yellow]{}{} & \Block[fill=green]{}{} & \Block[fill=yellow]{}{} & \Block[fill=green]{}{} & \Block[fill=yellow]{}{} & \Block[fill=green]{}{} & \Block[fill=yellow]{}{} & \Block[fill=green]{}{} & \Block[fill=yellow]{}{} & \Block[fill=green]{}{} & \Block[fill=yellow]{}{} & \Block[fill=green]{}{} & \Block[fill=yellow]{}{} & \Block[fill=green]{}{} & \Block[fill=yellow]{}{} & \Block[fill=green]{}{} & \Block[fill=yellow]{}{} & \Block[fill=green]{}{} \\ \hline
700   &   & \Block[fill=red]{}{} &   & \Block[fill=red]{}{} &   & \Block[fill=red]{}{} &   & \Block[fill=red]{}{} &   & \Block[fill=red]{}{} &    & \Block[fill=red]{}{} &    & \Block[fill=red]{}{} &    & \Block[fill=red]{}{} &    & \Block[fill=red]{}{} &    & \Block[fill=red]{}{} &    & \Block[fill=red]{}{} &    \\ \hline
1000  & \Block[fill=green]{}{} & \Block[fill=yellow]{}{} & \Block[fill=green]{}{} & \Block[fill=yellow]{}{} & \Block[fill=green]{}{} & \Block[fill=yellow]{}{} & \Block[fill=green]{}{} & \Block[fill=yellow]{}{} & \Block[fill=green]{}{} & \Block[fill=yellow]{}{} & \Block[fill=green]{}{} & \Block[fill=yellow]{}{} & \Block[fill=green]{}{} & \Block[fill=yellow]{}{} & \Block[fill=green]{}{} & \Block[fill=yellow]{}{} & \Block[fill=green]{}{} & \Block[fill=yellow]{}{} & \Block[fill=green]{}{} & \Block[fill=yellow]{}{} & \Block[fill=green]{}{} & \Block[fill=yellow]{}{} & \Block[fill=green]{}{} \\ \hline
1500  &   & \Block[fill=red]{}{} &   & \Block[fill=red]{}{} &   & \Block[fill=red]{}{} &   & \Block[fill=red]{}{} &   & \Block[fill=red]{}{} &    & \Block[fill=red]{}{} &    & \Block[fill=red]{}{} &    & \Block[fill=red]{}{} &    & \Block[fill=red]{}{} &    & \Block[fill=red]{}{} &    & \Block[fill=red]{}{} &    \\ \hline
2000  & \Block[fill=green]{}{} & \Block[fill=yellow]{}{} & \Block[fill=green]{}{} & \Block[fill=yellow]{}{} & \Block[fill=green]{}{} & \Block[fill=yellow]{}{} & \Block[fill=green]{}{} & \Block[fill=yellow]{}{} & \Block[fill=green]{}{} & \Block[fill=yellow]{}{} & \Block[fill=green]{}{} & \Block[fill=yellow]{}{} & \Block[fill=green]{}{} & \Block[fill=yellow]{}{} & \Block[fill=green]{}{} & \Block[fill=yellow]{}{} & \Block[fill=green]{}{} & \Block[fill=yellow]{}{} & \Block[fill=green]{}{} & \Block[fill=yellow]{}{} & \Block[fill=green]{}{} & \Block[fill=yellow]{}{} & \Block[fill=green]{}{} \\ \hline
3000  &   & \Block[fill=red]{}{} &   & \Block[fill=red]{}{} &   & \Block[fill=red]{}{} &   & \Block[fill=red]{}{} &   & \Block[fill=red]{}{} &    & \Block[fill=red]{}{} &    & \Block[fill=red]{}{} &    & \Block[fill=red]{}{} &    & \Block[fill=red]{}{} &    & \Block[fill=red]{}{} &    & \Block[fill=red]{}{} &    \\ \hline
4000  & \Block[fill=green]{}{} & \Block[fill=yellow]{}{} & \Block[fill=green]{}{} & \Block[fill=yellow]{}{} & \Block[fill=green]{}{} & \Block[fill=yellow]{}{} & \Block[fill=green]{}{} & \Block[fill=yellow]{}{} & \Block[fill=green]{}{} & \Block[fill=yellow]{}{} & \Block[fill=green]{}{} & \Block[fill=yellow]{}{} & \Block[fill=green]{}{} & \Block[fill=yellow]{}{} & \Block[fill=green]{}{} & \Block[fill=yellow]{}{} & \Block[fill=green]{}{} & \Block[fill=yellow]{}{} & \Block[fill=green]{}{} & \Block[fill=yellow]{}{} & \Block[fill=green]{}{} & \Block[fill=yellow]{}{} & \Block[fill=green]{}{} \\ \hline
7000  &   & \Block[fill=red]{}{} &   & \Block[fill=red]{}{} &   & \Block[fill=red]{}{} &   & \Block[fill=red]{}{} &   & \Block[fill=red]{}{} &    & \Block[fill=red]{}{} &    & \Block[fill=red]{}{} &    & \Block[fill=red]{}{} &    & \Block[fill=red]{}{} &    & \Block[fill=red]{}{} &    & \Block[fill=red]{}{} &    \\ \hline
10000 & \Block[fill=green]{}{} & \Block[fill=yellow]{}{} & \Block[fill=green]{}{} & \Block[fill=yellow]{}{} & \Block[fill=green]{}{} & \Block[fill=yellow]{}{} & \Block[fill=green]{}{} & \Block[fill=yellow]{}{} & \Block[fill=green]{}{} & \Block[fill=yellow]{}{} & \Block[fill=green]{}{} & \Block[fill=yellow]{}{} & \Block[fill=green]{}{} & \Block[fill=yellow]{}{} & \Block[fill=green]{}{} & \Block[fill=yellow]{}{} & \Block[fill=green]{}{} & \Block[fill=yellow]{}{} & \Block[fill=green]{}{} & \Block[fill=yellow]{}{} & \Block[fill=green]{}{} & \Block[fill=yellow]{}{} & \Block[fill=green]{}{} \\ \hline
\end{NiceTabular}
\caption{Distribution of settling filament evolutions into training and testing data sets by elasto-gravitational number and initial angle of inclination. The training data set is denoted by the green cells, testing data set A by the yellow cells, and testing data set B by the red cells.} 
\label{DataDistribution}
\end{table*}

\begin{figure*}
\centering
\includegraphics[width=1\textwidth]{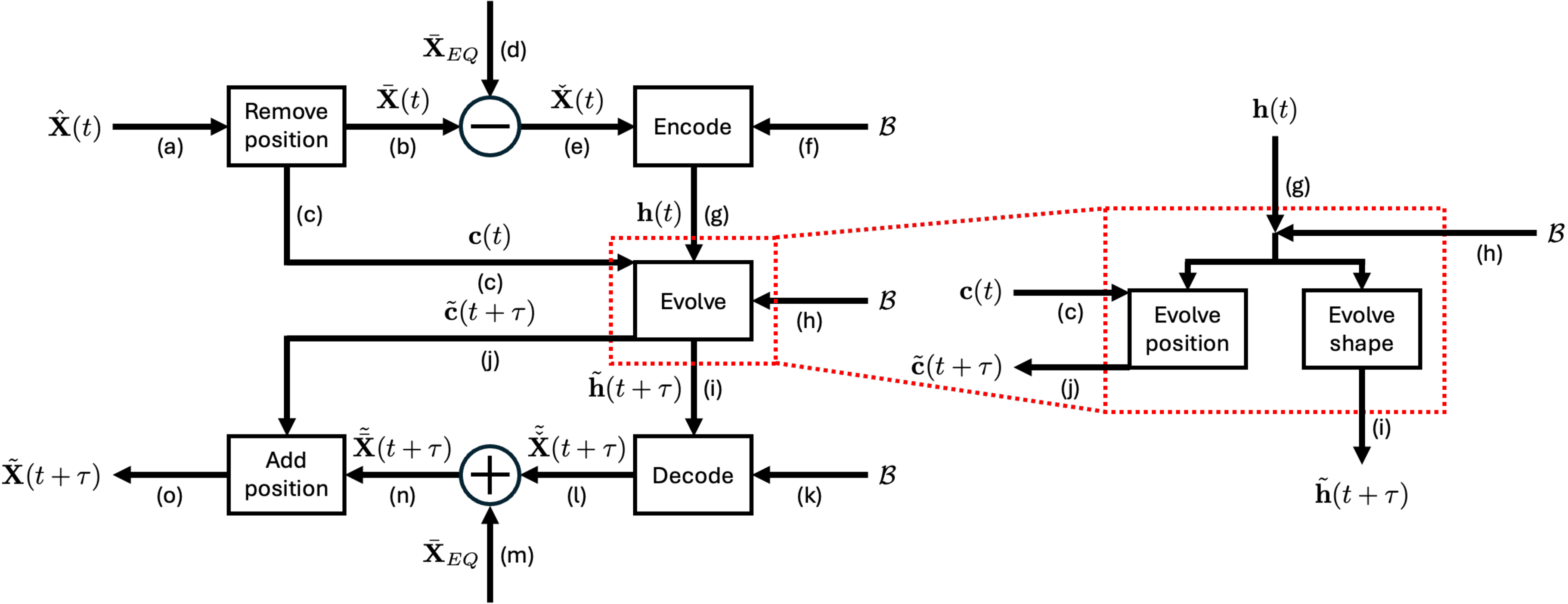}
\caption{Block diagram for data-driven model combining the autoencoder and temporal-evolution scheme. The temporal-evolution neural network, expanded in red, can be separated into two distinct neural networks forecasting the evolution of latent representation of the shape, $\mathbf{h}(t+\tau)$, and the shape-dependent change in position, $\mathbf{c}$; in practice, these can be forecasted by a single neural network.}
\label{NeuralNetworkDiagram}
\end{figure*}

A single model was developed to forecast the dynamics of the fiber across all $\mathcal{B}$ within the scope of this investigation.
A block diagram of the data-driven low-dimensional model architecture is shown in Figure \ref{NeuralNetworkDiagram}.
In the model, the position of the beads in a filament $\hat{\mathbf{X}}(t)$ [connection (a) in the block diagram], normalized by the half filament length $L/2$, is separated into the shape of the filament $\bar{\mathbf{X}}(t)$ [(b)], where the position of the center bead of the fiber subtracted from each bead position, and the position of center bead $\mathbf{c}(t)$ [(c)], such that $\bar{\mathbf{X}}(t) = \hat{\mathbf{X}}(t) - \mathbf{c}(t)$.
The terminal shape of the filament $\bar{\mathbf{X}}_{term} [(d)](\mathbf{B})$ (i.e. $\bar{\mathbf{X}}(t)$ as $t \rightarrow \infty$) is then subtracted from the filament shape to give the residual shape of the filament $\check{\mathbf{X}}$ [(e)].
An autoencoder neural network architecture, composed of an encoder and a decoder neural networks, learns a nonlinear coordinate transformation to reduce the dimensionality of the fiber shape to a low-dimension latent state and reconstruct the full shape from the latent state.
The encoder $\mathcal{E}(\check{\mathbf{X}}(t))$ first compresses the shape to a  low-dimensional latent state, $\mathbf{h}(t)$ [(g)], denoted as
\begin{equation}
\mathbf{h}(t)=\mathcal{E}(\check{\mathbf{X}}(t) ; \mathcal{B} ; \theta_E),
\end{equation}
where the latent dimension $d_h$ is much less than the ambient dimension $d_{\check{\mathbf{X}}} = 3N = 93$.
The decoder neural network, $\mathcal{D}(\mathbf{h}(t))$ then uses the low-dimensional latent state  to reconstruct the full state of the fiber shape, $\tilde{\check{\mathbf{X}}}(t)$, such that
\begin{equation}
\tilde{\check{\mathbf{X}}}(t)=\mathcal{D}(\mathbf{h}(t); \mathcal{B}; \theta_D)
\end{equation}

Both the encoder and decoder can receive the elasto-gravitational number as an additional input [(f) \& (k)], the necessity of which will be explored below.

The encoder and decoder networks are trained simultaneously, learning the neural network weights $\theta_E$ and $\theta_D$ using gradient descent to minimize the reconstruction error, given by
\begin{equation}
\mathcal{L}\left(\check{\mathbf{X}} ; \theta_E, \theta_D\right)=\left\langle\left\|\check{\mathbf{X}}-\mathcal{D}\left(\mathcal{E}\left(\check{\mathbf{X}}; \mathcal{B} ; \theta_E\right); \mathcal{B} ; \theta_D\right)\right\|_2^2\right\rangle,
\end{equation}
where  $\langle \cdot \rangle$ is the average over the training data.

A time-integrating neural network is then trained to use the current latent state and the parameter value $\mathcal{B}$ [(h)] to forecast the evolution of the latent space one time step $\tau$ into the future, $\mathbf{h}(t+\tau)$ [(i)], while simultaneously forecasting the new position, $\mathbf{c}(t+\tau)$ [(j)].

In this study, we perform time-integration though the use of a neural ordinary differential equation (ODE), a neural network trained to forecast the right-hand side of the temporal evolution of the latent state, given as
\begin{equation}
\frac{d \mathbf{h}}{d t}=\mathcal{F}_h(\mathbf{h}(t); \mathcal{B}; \theta_h)-A \mathbf{h}.\label{eq:NODEh}
\end{equation}
The final term on the right-hand side provides an explicit damping, which has been shown to stabilize the latent state dynamics for neural ODEs\citep{Linot2023}. 
The evolution of the center bead position can be similarly modeled by a neural ordinary differential equation, wherein the neural network reconstructs the right-hand side of 
\begin{equation}
\frac{d \mathbf{c}}{d t}=\mathcal{F}_c(\mathbf{h}(t); \mathcal{B}; \theta_c),\label{eq:NODEc}
\end{equation}
\noindent where $\mathbf{c}(0) = \mathbf{0}$. It should be noted that while both the change in latent shape and position depend on the current latent shape of the fiber, both are independent of the current position, allowing the model to forecast the evolution of the filament at any arbitrary position.

The latent state and center bead positional change are then integrated forward in time to forecast the temporal evolution of the latent state.
The latent state one time step forward, $\mathbf{h}(t+\tau)$, can then be calculated as 

\begin{equation}
\tilde{\mathbf{h}}(t+\tau)=\mathbf{h}(t)+\int_t^{t+\tau}\left(\mathcal{F}_h\left(\mathbf{h}\left(t^{\prime}\right); \mathcal{B} ; \theta_h\right)-A \mathbf{h}\right) d t^{\prime},
\end{equation}

\noindent and similarly the position one time step forward, $\mathbf{c} (t+\tau)$, as 

\begin{equation}
\tilde{\mathbf{c}}(t+\tau)= \mathbf{c}(t)+\int_t^{t+\tau}\mathcal{F}_c\left(\mathbf{h}\left(t^{\prime}\right); \mathcal{B} ; \theta_F\right) d t^{\prime}.
\end{equation}

In practice, the evolution of the latent state and positional change can be forecasted by a single neural ODE, $\mathcal{F}(\bar{\mathbf{h}}(t); \mathcal{B}; \theta_F)$, where the latent state and the  position vector  are concatenated as $\bar{\mathbf{h}}(t) = [\mathbf{h}(t), \mathbf{c}(t)]$.
The time-integrating neural network is trained to learn the neural network weights $\theta_F$ using a standard stochastic gradient descent to minimize the forecasting error, with a loss given by

\begin{equation}
\mathcal{L}\left(\mathbf{h} ; \theta_F\right)=\left\langle\left\|\mathbf{h}(t+\tau)-\tilde{\mathbf{h}}(t+\tau)\right\|_2^2\right\rangle + \left\langle\left\| \mathbf{c} (t+\tau)-\tilde{\mathbf{c} }(t+\tau)\right\|_2^2\right\rangle.
\end{equation}

The two components of the loss are given equal weighting, as the relative magnitudes of each term are sufficiently comparable that neither dominates in training. The resulting forecast of the latent state can then be decoded to produce a predicted latent shape $\tilde{\check{\mathbf{X}}}(t+\tau)$ [(l)] and combined with the terminal shape filament shape [(m)] to recover the forecast of the filament shape $\tilde{\bar{\mathbf{X}}}(t+\tau)$ [(n)].
The forecasts of the shape and position can then be recombined to produce a full forecast of the evolution of the settling fiber $\tilde{\hat{\mathbf{X}}}(t+\tau)$ [(o)].
By time-integrating the neural ODE from a given initial condition, the dynamics of the full filament can be modeled in the low-dimensional state until the terminal position is reached.

\section{Results and discussion}

\begin{figure*}
\centering
\includegraphics[width=1\textwidth]{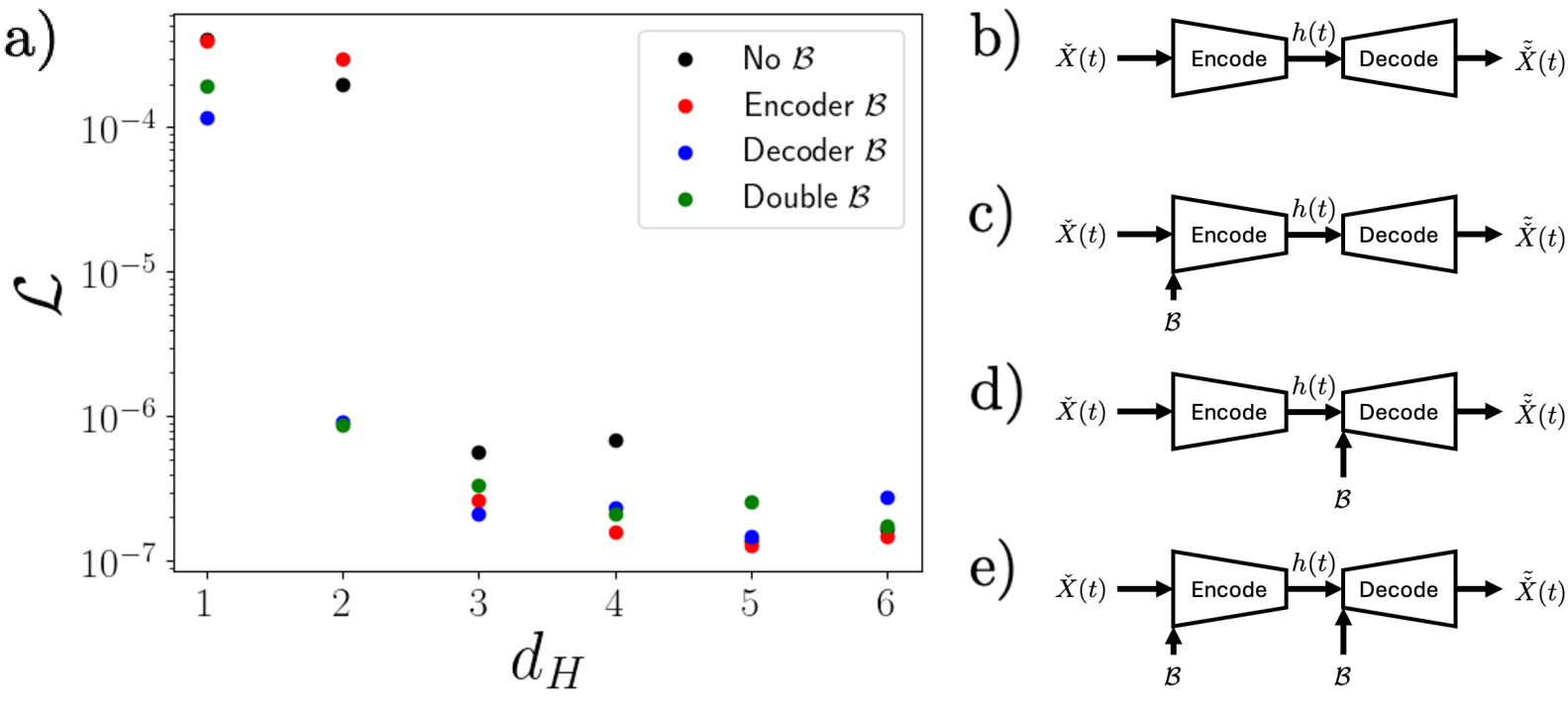}
\caption{a) Loss over the testing data for each autoencoder architecture as a function of latent dimension. Block diagrams for the autoencoder neural network architectures (b) ``No $\mathcal{B}$'', (c) ``Encoder $\mathcal{B}$'', (d) ``Decoder $\mathcal{B}$'', and (e) ``Double $\mathcal{B}$''.}
\label{DimensionReduction}
\end{figure*}

We first sought to determine the necessary autoencoder architecture with which to model the evolution of a filament at multiple elasto-gravitational numbers.
We evaluated the performances of four distinct architectures.
The first architecture, denoted as ``No $\mathcal{B}$'' did not receive the elasto-gravitational number in either the encoder or the decoder, such that $\tilde{\check{\mathbf{X}}} = D\left(E\left(\check{\mathbf{X}} ; \theta_E\right) ; \theta_D\right)$ and connections (f) and (k) are removed.
The second, referred to as ``Encoder $\mathcal{B}$'', received the elasto-gravitational number as an input solely to the encoder, such that $\tilde{\check{\mathbf{X}}} = D\left(E\left(\check{\mathbf{X}}; \mathcal{B} ; \theta_E\right) ; \theta_D\right)$ and only connection (f) is removed.
Similarly, the third architecture, referred to as ``Decoder $\mathcal{B}$'', received the elasto-gravitational number as an input solely to the decoder such that $\tilde{\check{\mathbf{X}}} = D\left(E\left(\check{\mathbf{X}}; \theta_E\right); \mathcal{B} ; \theta_D\right)$ and only connection (f) is removed.
The final architecture, denoted as ``Double $\mathcal{B}$'' received the elasto-gravitational number as an input to both the encoder and the decoder, such that $\tilde{\check{\mathbf{X}}} = D\left(E\left(\check{\mathbf{X}}; \mathcal{B} ; \theta_E\right); \mathcal{B} ; \theta_D\right)$ and both connections (f) and (k) are included.
Block diagrams of these architectures are shown in Figure \ref{DimensionReduction}.b-e.
Each architecture was used to train autoencoders at latent dimensions $1 \leq d_h \leq 6$ with the training data set.
The reconstruction loss across all of the testing data was calculated for the models at each latent dimension, and the accuracy of the coordinate transformations learned by the encoders and decoders evaluated.

Shown in Figure \ref{DimensionReduction} is the reconstruction loss for test data sets A and B at each latent dimension.
The performance of each autoencoder architecture is similar, with the loss dropping precipitously at a latent dimension between two and four, with higher latent dimensions showing no significant improvement in the reconstruction loss.
A latent dimension of four was selected as the standard size for the remainder of this study, to ensure maximum accuracy while still achieving substantial dimension reduction as $d_h = 4 \ll d_{\check{\mathbf{X}}} = 93$, with an additional three degrees of freedom needed to account for filament position.
Of note, none of the autoencoder architectures resulted in significantly lower reconstruction losses, indicating that any of the latent state parametrizations can be used to develop a dynamical model.
By heuristic observations, the ``Decoder $\mathcal{B}$'' was observed to produce the most consistent results, and as such was selected for continued use in this study, although it should be noted that all architectures can produce accurate models.

Once the appropriate autoencoder architecture and required number of degrees of freedom were  determined, the data-driven model utilizing neural ODEs was applied to forecast the evolution of all filament evolutions in test data sets A and B. as described in table \ref{DataDistribution}.
The initial bead positions of the filament in each trajectory were input to the model and the shape and positional change forecasted using Eqs.~\ref{eq:NODEh}-\ref{eq:NODEc} over the settling period, $T$, which is dependent on the initial orientation and elasto-gravitational number.
The time-dependent error $E(t)$ for each forecast was calculated as
\begin{equation}
E(t) = \frac{\left\langle\left\|\bar{\mathbf{X}}(t)-\tilde{\bar{\mathbf{X}}}(t)\right\|_2^2\right\rangle} {D_{shape}} + \frac{\left\langle\left\|\bar{\mathbf{c}}(t)-\tilde{\bar{\mathbf{c}}}(t)\right\|_2^2\right\rangle} {D_{pos}}.
\end{equation}
The first term in the error formula calculates the error in the forecasted shape of the filament and is normalized by the error between two filaments offset by one filament length, $D_{shape} = \|\bar{\mathbf{X}}-(\bar{\mathbf{X}}+L)\|_2^2$.
The second term calculates the error in the forecasted position of the center bead, normalized by the total distance covered by the true data, $D_{pos} = \left\|\bar{\mathbf{c}}(t)-\bar{\mathbf{c}}(0)\right\|_2^2$.
The best and worst forecasts at each $\mathcal{B}$ were identified by selecting the forecasts with the lowest and highest maximum instantaneous error, respectively.
Here, we will show representations of the forecasts of these selected trajectories, and the accuracy of the remaining trajectories within the test data sets can be inferred to fall between the bounding forecasts shown.
Videos of all forecasted trajectories in the test data sets can be found in the Supplemental Material for further elucidation\citep{SupplementaryVideos}.

\begin{figure*}
\centering
\includegraphics[width=1\textwidth]{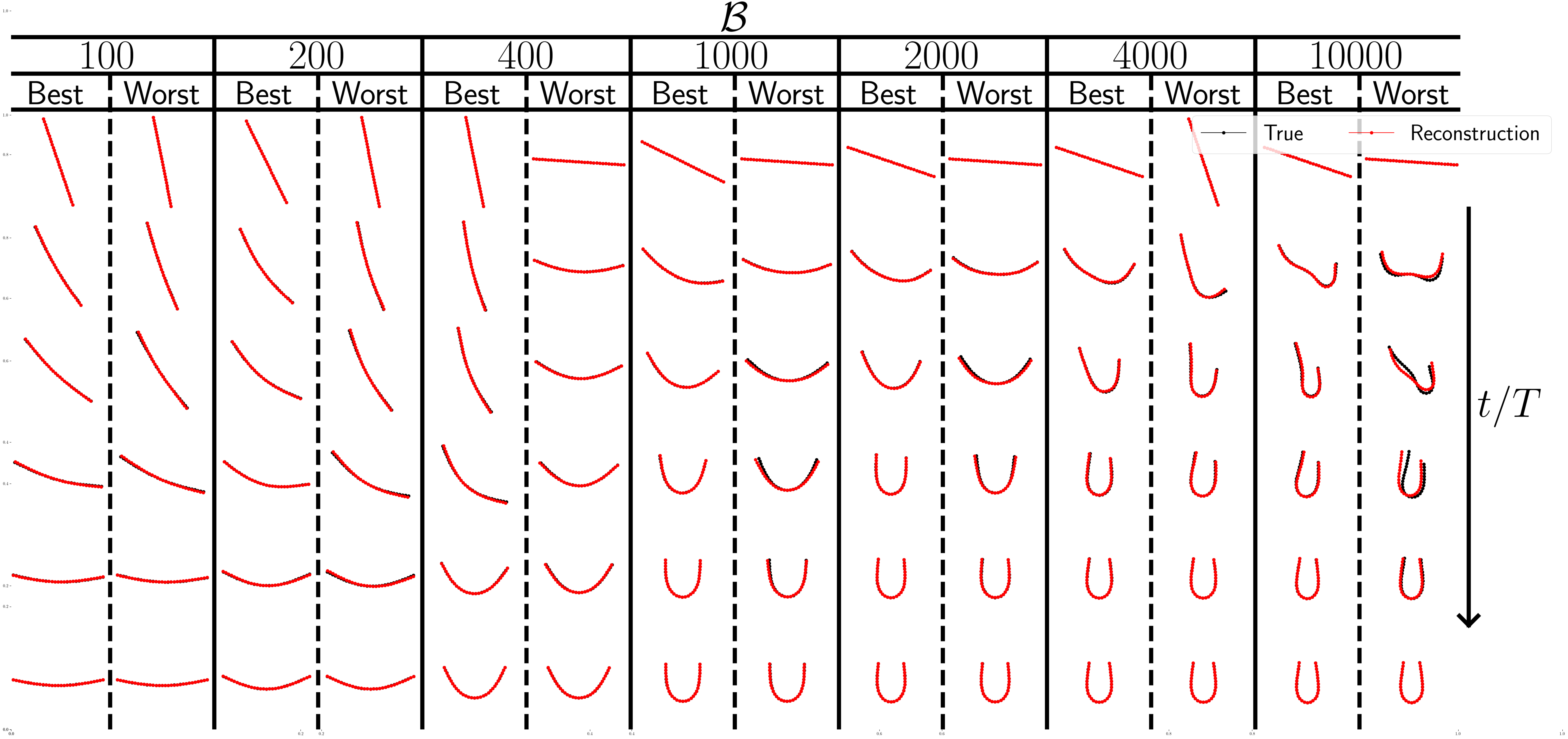}
\caption{Evolution of the shape of a filament settling in quiescent Newtonian fluid given an initial orientation from the best and worst forecasts (red) and the true evolution (black) at a given $\mathcal{B}$. Here, $\mathcal{B}$ is within the training data set, and the initial angles of inclination are not.}
\label{ShapeComp_B=in}
\end{figure*}

We first compared the predicted evolution of the shape of the filament over the settling period to the true trajectory.
Shown in Figure \ref{ShapeComp_B=in} are snapshots of the shape of the filaments from the best and worst forecasts for reconstructed trajectories in test data set A over the settling period, overlaying the true shape of the filament at that time.
For all trajectories, the predicted shape of the filament closely matches the true shape across the settling period, showing our model has successfully interpolated the shape evolution to new initial angles of orientation.
The best forecasts across test data set A show little deviation from the true evolutions, with the underlying true shapes nearly indistinguishable from the predictions.
The terminal shapes of the forecasts all trend towards the true terminal shape, demonstrating that the model has learned the appropriate long time behavior of the residual shape.
The worst forecast of the shape over this test data set occurs at $\mathcal{B} = 10000$, where deviations between the true and reconstructed shape are apparent; this error is caused by a slight lag in the forecasted reorientation of the filament, resulting in a time-delay between the instantaneous shapes.
It should be noted that the initial orientations of the best and worst forecasts are reasonably evenly distributed across the initial angles of inclinations, showing that our data-driven model performs equally well across the initial angle distribution and does not favor any obvious subset of inclinations.

\begin{figure*}
\centering
\includegraphics[width=1\textwidth]{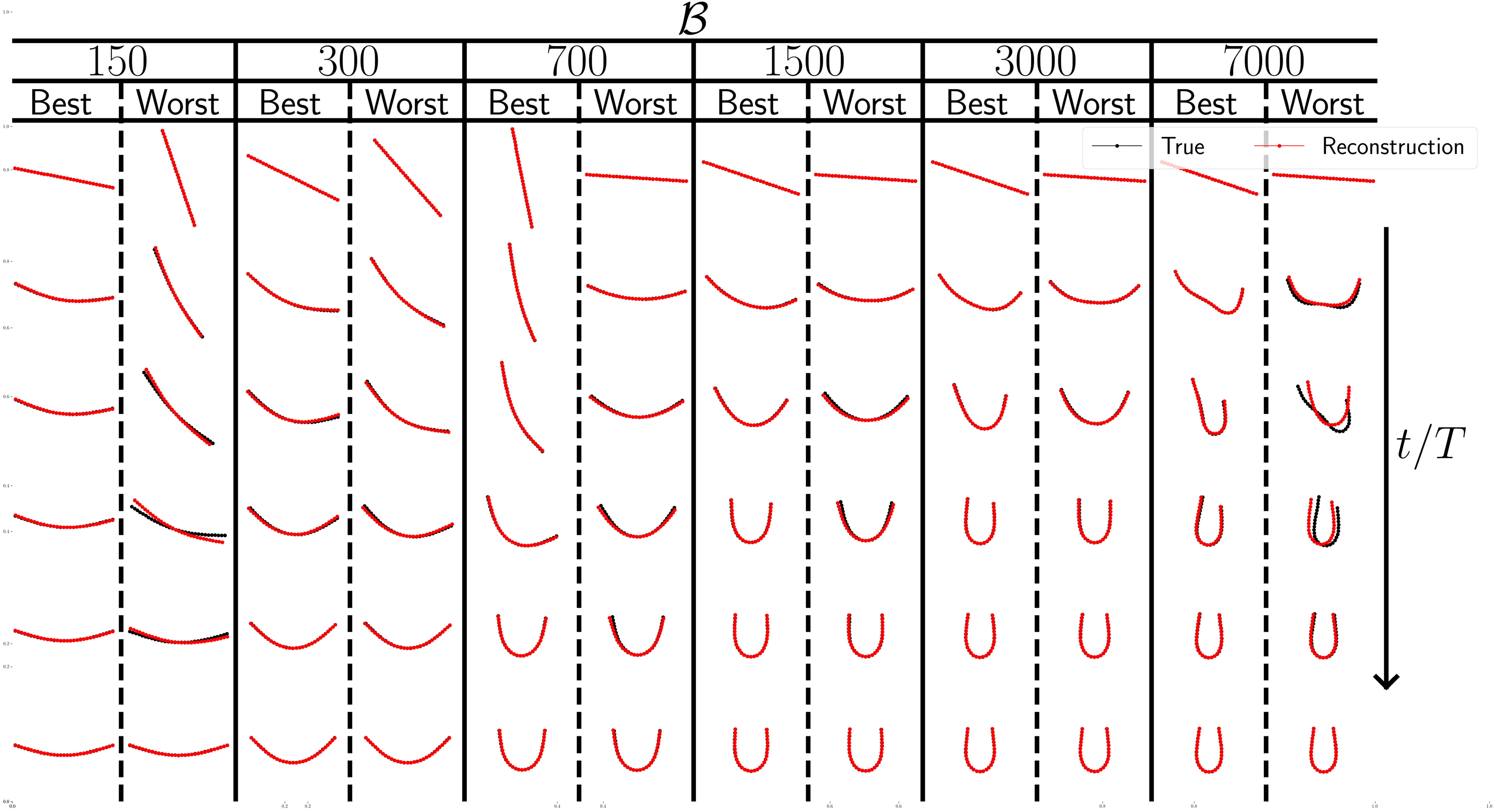}
\caption{Evolution of the shape of a filament settling in quiescent Newtonian fluid given an initial orientation from the best and worst forecasts (red) and the true evolution (black) at a given $\mathcal{B}$. Here, neither $\mathcal{B}$ nor the initial angles of inclination are within the training data set.}
\label{ShapeComp_B=out}
\end{figure*}

The predicted shapes of the filaments over the settling period in test data set B were then compared to the true shapes from the trajectories, shown in \ref{ShapeComp_B=out} with the reconstructed shapes overlaid on the true from the best and worst forecast at each $\mathcal{B}$.
As with the forecasts from test data set A, the forecasted shapes here closely match the true shapes, with only minor deviations observed across the test data set, demonstrating that our model has successfully interpolated the shape evolution to new elasto-gravitational numbers.
The reconstructed shapes from the best forecasts almost completely match the true shapes, with only small errors between the shapes apparent.
Again, the predicted terminal shapes are all indistinguishable from the true, indicating that our model has correctly identified the long time collapse of the residual shape at untrained $\mathcal{B}$-dependent terminal shapes.
The worst forecasts in test data set B, here at $\mathcal{B} =$ 150 \& 7000, show noticeable deviations between the true and reconstructed shapes at intermediate times, due again to a lag in the start of the reorientation behavior between the two trajectories.
Once more, the initial angle of orientations for the best and worst forecasts show no grouping towards common $\theta_0$, meaning that, at these interpolated $\mathcal{B}$, the model still does not show preference towards forecasts of any discernible subset of inclination angles.

\begin{figure*}
\centering
\includegraphics[width=1\textwidth]{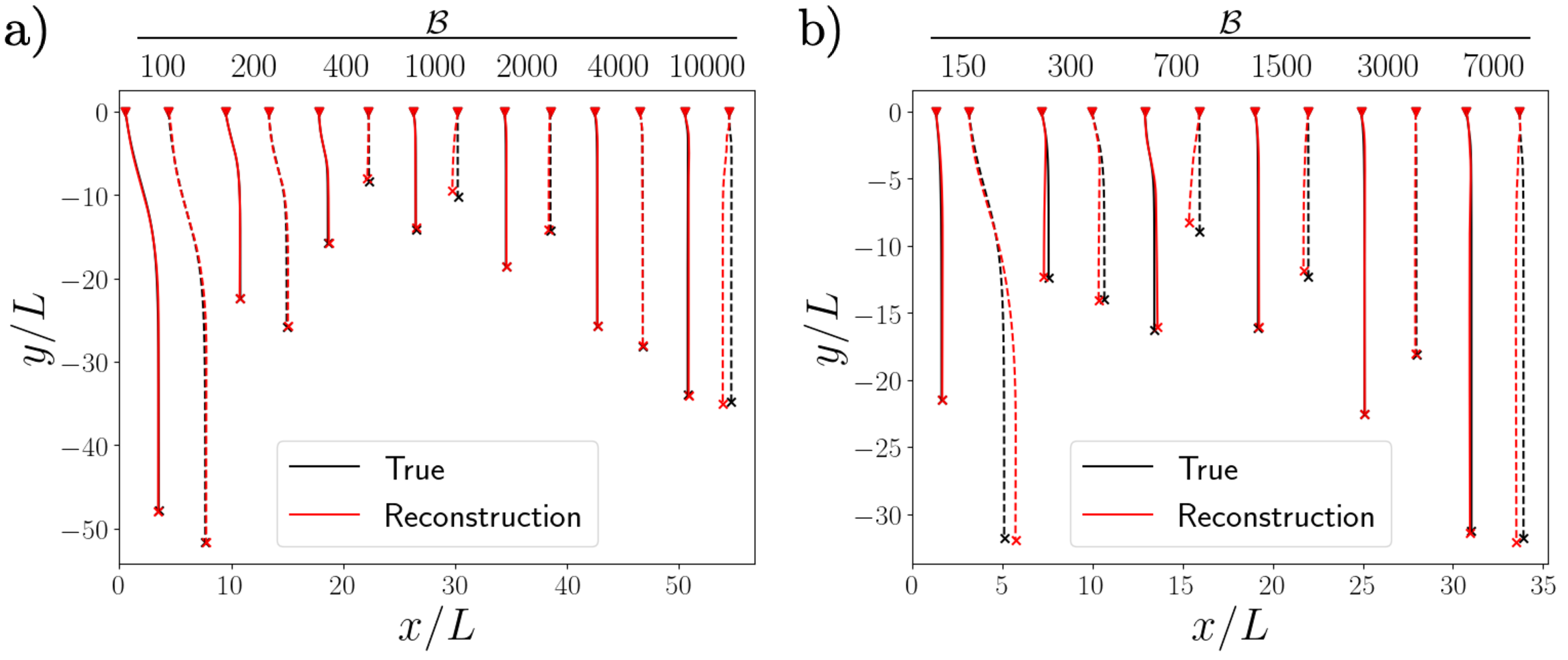}
\caption{Trajectory of the center bead $c(t)$ a filament settling  in quiescent Newtonian fluid given an initial orientation from the best (solid lines) and worst (dashed lines) forecasts and the true evolution at a given $\mathcal{B}$. The initial positions are denoted by symbol “$\blacktriangledown$” and the terminal positions are denoted by the symbol “$\times$”, with the true and predicted positions in black and red, respectively. In (a), the $\mathcal{B}$ are within the training data set, but at initial angles of inclination not within the training data; in (b),  both $\mathcal{B}$ and the initial angles of inclination are not within the training data.}
\label{TrajectoryComp}
\end{figure*}

We next investigated the accuracy of the forecasts of the center bead trajectory by our data-driven model.
Shown in Figure \ref{TrajectoryComp} is a comparison between the true and forecasted trajectories of the center bead $\mathbf{c}(t)$ in the filament evolution from test data set A in \ref{TrajectoryComp}.a and from test data set B in \ref{TrajectoryComp}.b, with the trajectories from best and worst forecasts at each $\mathcal{B}$ on the left and right, respectively. 
As the initial position of the filament is arbitrary, we have set all trajectories to begin on the $x$-axis, offset in the $x$-direction for visual clarity.
As can be seen, the forecasted positional evolution of the center bead closely tracks the true positions for all predicted trajectories.
The best forecasts of the trajectories at each $\mathcal{B}$ in both data sets are nearly identical to the true trajectories, with only slight positional errors between them; many of the trajectories from the worst forecasts across the test data sets show a similar degree of accuracy to the true positions.
In some of the worst forecasts, the reconstructed trajectories are appreciably different from the true; however, in contrast to the total distance traversed by the center bead, the error between the true and forecasted positions is comparatively  small across the calculated trajectories.

\begin{figure*}
\centering
\includegraphics[width=1\textwidth]{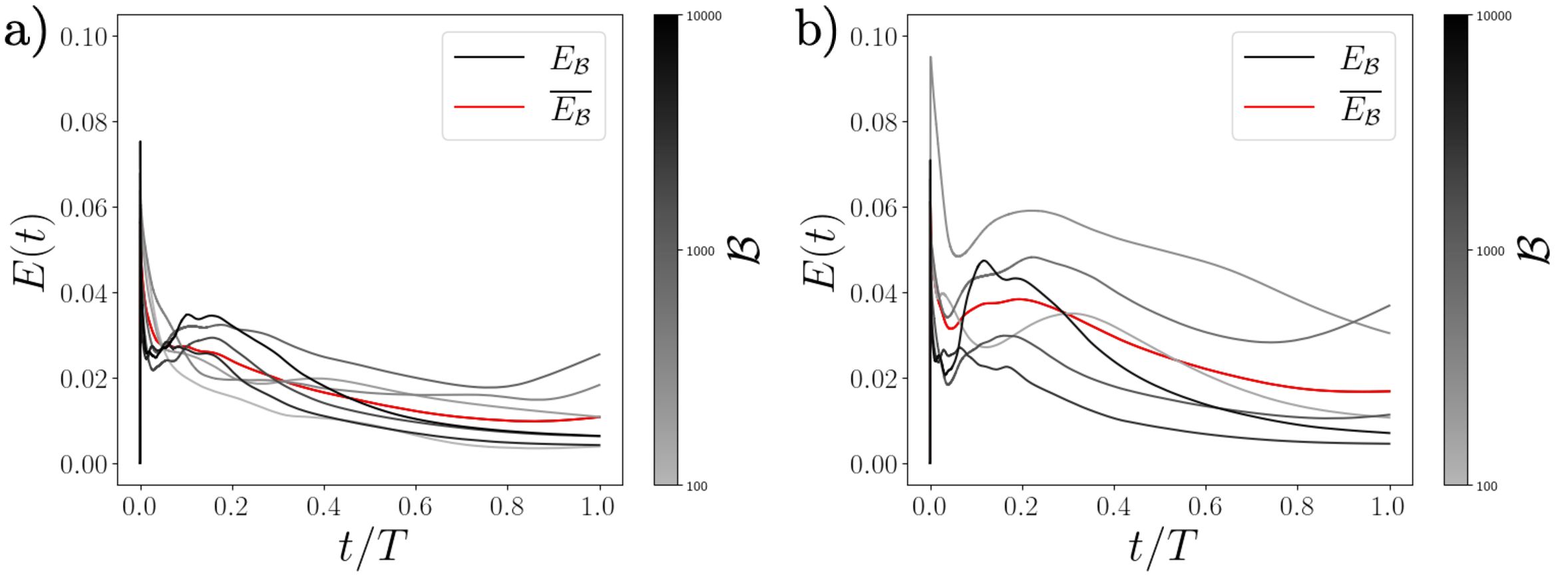}
\caption{Ensemble-average error vs. time for forecasts of the evolution of a filament settling in quiescent Newtonian fluid at each $\mathcal{B}$ (gray) and averaged over all $\mathcal{B}$ (red). In (a), the $\mathcal{B}$ are within the training data set, but at initial angles of inclination not within the training data; in (b),  both $\mathcal{B}$ and the initial angles of inclination are not within the training data.}
\label{Errors}
\end{figure*}

Finally, we investigated the error in the forecast across the investigated elasto-gravitational numbers in both test data sets.
In Figure \ref{Errors}, the ensemble-averaged error over the settling period is shown at each $\mathcal{B}$ for test data set A in \ref{Errors}.a and for test data set B in \ref{Errors}.b, as well as the ensemble-averaged error across the $\mathcal{B}$ in each data set.
At short times, the instantaneous error shows a sudden increase associated with the positional error, due to small total distance covered compared to the initial error between the change in position for the true and reconstructed trajectories; as time  increases, the error associated with the positional error decreases precipitously as the total distance covered grows much larger than the slight difference in position.
At intermediate times, lags between the true and forecasted reorientation behaviors cause instantaneous error due to difference between the shapes, causing an increase in error in a limited number of trajectories.
At long times, the error trends toward a small but finite value, as slight differences between forecasted terminal shape and terminal velocity cause an unavoidable deviation between the trajectories.
At all $\mathcal{B}$ in both test data sets, the error remains small for the entire settling period, with each ensemble-averaged error remaining below 0.1 in all cases.

\section{Conclusions}

In this work, we have developed a data-driven modeling technique for learning a low-dimensional model for the dynamics of an elastic fiber at arbitrary initial angle of inclination settling under gravity in a quiescent, viscous Newtonian fluid.
We have demonstrated that our proposed neural network architecture can capture that filament shape in a latent state with significantly fewer dimensions than the full state space, and we have shown that the architecture needed to adequately reduce the dimension is aided by, but does not require, the elasto-gravitational number of the filament as an additional input.
We have shown that the model can forecast the evolution of filaments at untrained angles of inclination, and can interpolate the evolution behavior to trajectories of fibers at unseen $\mathcal{B}$.
The forecasts of the shape of the settling fiber closely match the true shapes across the settling period for all investigated $\mathcal{B}$, with only small deviations due to a lag in reorientation time.
The predicted trajectories of the center bead position closely match the true trajectories, with the positional error remaining small in comparison to the total distance covered by the filament.
We have shown that, across all trajectories, in both test data sets, the instantaneous ensemble-averaged error remains low for the entire settling period.

In this study, we have developed a technique to model the dynamics of a flexible particle, and we have demonstrated that our proposed model is able to accurately forecast the evolution of a simple elastic structure.
It remains an open question as to the optimal architecture with which to model a settling filament.
In this paper, we developed our model at a latent dimension that, while significantly lower than the ambient dimension of the system, is not known to be the minimal dimension of this system.
Future study can reveal if, given improved modeling techniques, the necessary latent dimension for this system can decrease \citep{Zeng2024}.
Additionally, changing the type of neural network used in the model architecture may lead to more accurate, lower dimensional models.
Here, we only used fully-connected, feed-forward dense neural networks in our data-driven models.
Given the localized nature of the data of the structure of an elastic particle, using an architecture that utilizes more localized information processing may produce better models, especially for objects with more geometric complexity.
Recent studies have shown that graph neural networks have been shown to improve models of multiphase flows\citep{Ma2024, Ma2022}. 
Given the inherent graph structure of a discretized flexible particle, graph neural networks could leverage this structure to improve model forecasts.

Now that we have demonstrated that our technique can model a simple elastic particle in a simple flow, future investigation should determine the ability of such models to forecast the dynamics of more complicated particles in differing flow fields.
For the flexible filament, this can include modeling the dynamics of such particles in flow fields such as shear flow and extensional flow.
Previous studies have examined the behavior of filaments in these flow fields using numerical analysis and physics-based simulations\citep{Słowicka2012, Słowicka2013, Słowicka2015, Farutin2016}. 
This can then be extended to more complex particles, such as elastic sheets and deformable spheres, as well as investigating particles with varying shapes at rest.
This can include developing low-dimensional models for biological systems, where cells suspended in flow behave as elastic particles.
In the long term, such models could be expanded to forecast the evolution of multiple elastic particles entrained in flow, capturing the inter-particle hydrodynamic interactions with a single, low-dimensional model. 

\nocite{*}

\end{document}